# Hep Cluster – First Step Towards Grid Computing
# 7th JK Science Congress
# Oct 13-15, 2011

**Vivek Chalotra\*, Anju Bhasin, Anik Gupta, Sanjeev Singh Sambyal**

Department of Physics, Baba Saheb Ambedkar Road, University of Jammu, Jammu-180006, India.
\*Email ID: vivekathep@gmail.com

## ABSTRACT

HEP Cluster is designed and implemented in Scientific Linux Cern 5.5 to grant High Energy Physics researchers one place where they can go to undertake a particular task or to provide a parallel processing architecture in which CPU resources are shared across a network and all machines function as one large supercomputer. It gives physicists a facility to access computers and data, transparently, without having to consider location, operating system, account administration, and other details. By using this facility researchers can process their jobs much faster than the stand alone desktop systems.

## INTRODUCTION

The basic unit of a cluster is a single computer, also called a "node". Clusters can grow in size - they "scale" - by adding more machines. A cluster as a whole will be more powerful and faster than the individual computers and their connection speeds. In addition, the operating system of the cluster must make the best use of the available hardware in response to changing conditions. This becomes more of a challenge if the cluster is composed of different hardware types (a "heterogeneous" cluster), if the configuration of the cluster changes unpredictably (machines joining and leaving the cluster), and the loads cannot be predicted ahead of time.

## OVERVIEW OF THE CLUSTER

In this cluster, there is one master node connected with external storage and three worker nodes. Researchers can login into the master node using the usernames which have been allotted to them and can select any worker node in which they want to perform their operations. They can split their jobs between different working nodes. After the completion of jobs, the results will automatically get stored in the centralized storage area. These results can be then analyzed by the user for the research purpose. In this way they can do their research jobs much faster than the individual computers. Fig 1.0 shows the overview of the cluster.

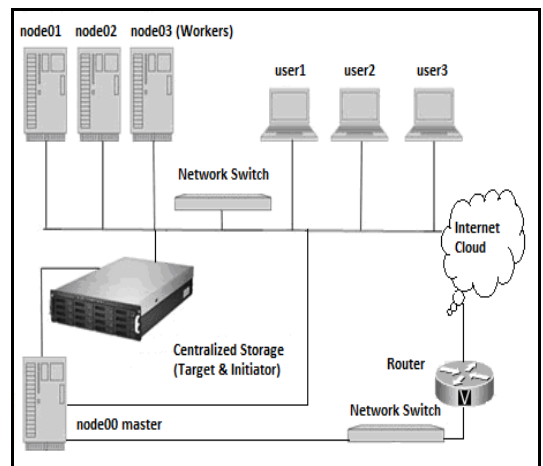

Fig. 1.0    Overview of the Cluster

## EQUIPMENT USED

**Hardware :** HP Proliant ML150 Servers, 100 Mbps internet lease line link provided by Ernet India(ISP), Cisco 1841 series router, Cisco Catalyst 2950 series and D-Link DES - 1008D switches, 42U rack and 5KVA UPS with 2 hours battery backup.

**Software:** Scientific Linux Cern release 5.5 and HEP research applications (Root, Aliroot & Geant).

## NETWORK SETUP

| Config | Master | Worker1 | Worker2 | Worker3 |
|---|---|---|---|---|
| H.N | node00 | node01 | node02 | node03 |
| Eth0 IP | 10.x.x.1 | 10.x.x.2 | 10.x.x.3 | 10.x.x.3 |
| Eth1 IP | 144.x.x.x | - | - | - |

## PASSWORD FREE SSH LOGIN

Edit the /etc/profile of all the nodes. Then copy the generated authorized_keys of all the workers in root/.ssh/authorized_keys file of the master and finally copy .ssh/authorized_keys file of the master in all the worker nodes to make all the cluster nodes password free among each other.

## INSTALLATION OF NODES

- Configure RAID 3 in all nodes.
- Partition table for the worker nodes(Using LVM Technique):
  /boot = 500MB
  /home = 100GB
  / = 100GB
  swap = 8GB
- Partition table for the master node(Using LVM Technique):
  /boot = 500MB
  /home = 100GB
  / = 100GB
  swap = 16GB
  /Jugrid = 4TB  {Storage}
- Install Scientific Linux Cern 5.5 in all the cluster nodes.

## NFS MOUNT CENTRALIZED STORAGE

To export /Jugrid of node00 to node01, node02 & node03 enter the following in /etc/exports file and restart the NFS service :
/jugrid node01(rw,rsync) node02(rw, rsync) node03(rw, rsync)
Now mount the centralized storage from the workers like the following:
# mount node00:/Jugrid /home
# echo "10.1.3.193:/Jugrid /home nfs defaults 0 0" >> /etc/fstab

## CREATE USERS

Create some users in /Jugrid using the below command and copy the /etc/passwd, /etc/group & /etc/shadow files of the master into the workers:
useradd -gusers -Gusers -s/bin/shell -pxxxx -d/Jugrid/vivek -m user1

## DISK QUOTA MANAGEMENT

Edit /etc/fstab file of node00 and add the following entry:
/dev/VolGroup00/LogVol00     /Jugrid   ext3   defaults,usrquota     1 2

# mount –o remount /Jugrid
#quotacheck –a /Jugrid
#quotaon /Jugrid
#edquota –u username

## INSTALLATION OF HEP RESEARCH APPLICATIONS IN /JUGRID (CENTRALIZED STORAGE)

Before the installation of Root, AliRoot and Geant in the centralized storage (/Jugrid) we have to set some environment variables in /etc/bashrc file of the master node in the cluster like the following:-

# Root environment
export ROOTSYS=/Jugrid/alice/root
export PATH=$ROOTSYS/bin\:$PATH
export LD_LIBRARY_PATH=$ROOTSYS/lib\:$LD_LIBRARY_PATH

# AliRoot environment
export ALICE=/Jugrid/alice
export ALICE_LEVEL=AliRoot
export ALICE_ROOT=$ALICE/$ALICE_LEVEL
export ALICE_TARGET=`root-config --arch`
export PATH=$ALICE_ROOT/bin/tgt_$ALICE_TARGET\:$PATH
export LD_LIBRARY_PATH=$ALICE_ROOT/lib/tgt_$ALICE_TARGET\:$LD_LIBRARY_PATH

# Geant3 environment
export PLATFORM=`root-config --arch`
export LD_LIBRARY_PATH=$ALICE/geant3/lib/tgt_$PLATFORM\:$LD_LIBRARY_PATH

**Root Installation:-**
mkdir /Jugrid/alice
cd /Jugrid/alice
svn co https://root.cern.ch/svn/root/tags/v5-26-00b root_v5.26.00
ln -s root_v5.26.00 root
cd $ROOTSYS
./configure
make
make install
After giving the command **make install** it will take 2 or more hours to complete the installation depending upon the speed of the machine. Then give **root** command to run the application.

**Geant3 Installation:-**
cd $ALICE
svn co https://root.cern.ch/svn/geant3/tags/v1-11 geant3
cd $ALICE/geant3
make

Please note that GEANT 3 is downloaded in **$ALICE** directory. This feature is used in AliRoot to set correctly the include path to TGeant3.h. Another important feature is the **PLATFORM** environment variable. If it is not set, the make file sets it to the result of `**root-config --arch**`.

**Aliroot Installation:-**
svn co https://alisoft.cern.ch/AliRoot/branches/v4-18-Release AliRoot_v4.18-Release
ln -s AliRoot_v4.18-Release AliRoot
cd $ALICE_ROOT
make
Installation complete.
Give **aliroot** command to run the application.

### CREATING ALIASES

We have installed root and aliroot in master node but we want users to run applications on the worker nodes rather than on master node. For that we have to create aliases for the commands root and aliroot like the following:-
[root@cluster Jugrid]# alias root='echo "Please login to any worker node from node01-node03 to run root/aliroot" '
[root@cluster Jugrid]# alias aliroot= 'echo "Please login to any worker node from node01-node03 to run root/aliroot" '

After creating the aliases, whenever a user tries to run root or aliroot on master node, he/she will get the following message:
 **"Please login to any working node from node01-node03 to run root/aliroot".**

### LOGIN SCREEN SETUP

Create a new ssh login message. Edit the following file of the master:
pico /etc/motd
Now type in the security message you wish all users to see once they login to master server through ssh:
*** WELCOME TO  HEP CLUSTER************
*******************************************
This facility is for authorized users only. All activity is logged and regularly checked by the administrator. From here you can login into any computing node from node01-node03 using the command ssh –Y node0x and then run root, aliroot & geant.
For any queries contact:-
Vivek Chalotra (vivekathep@gmail.com)
******************************************
******************************************

### MONITORING

Use the iptraf utility for monitoring incoming & outgoing traffic of the cluster. Fig 1.2 shows the iptraf utility.

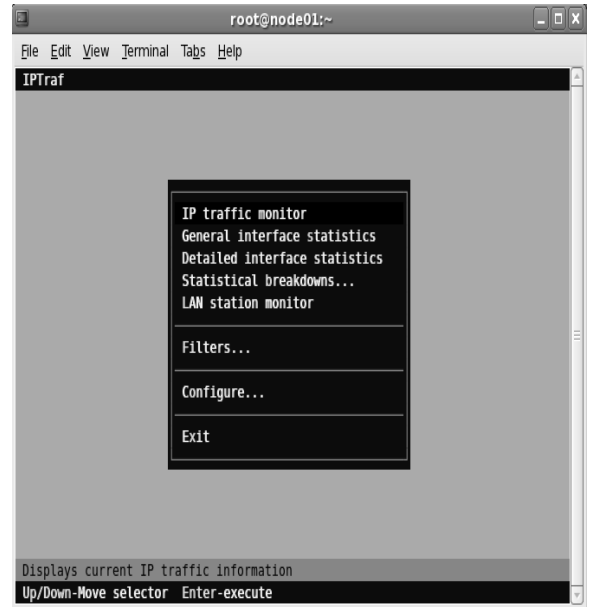

Fig 1.1   IPTRAF

### STARTING AND STOPPING SEQUENCE

**Starting sequence of the cluster:**
1. Switch on the UPS
2. Power on the master node.
3. Power on the worker nodes.

**Stopping sequence of the cluster:**
1. Shutdown the worker nodes.
2. Shutdown the master node.
3. **Switch off the UPS.**

**Note: - Do not switch off the router and switches.**